\documentstyle[12pt,epsf]{article}
\input epsf

\newcount\xrefpos \xrefpos=0
\newcount\yrefpos \yrefpos=0
\newcount\xput \xput=0
\newcount\yput \yput=0
\def\refpos#1 #2 #3{\global\xrefpos=#1 \global\yrefpos=#2
                         \rlap{$\smash{#3}$}}
\def\put #1 #2 #3{\xput=#1 \yput=#2
                  \advance\xput by -\xrefpos
                  \advance\yput by -\yrefpos
                  \rlap{\kern\the\xput truebp
                        \vbox to 0pt{\vss\hbox{$\displaystyle #3$}%
                        \kern\the\yput truebp}}}
\def\beginlabels\refpos#1\endlabels{\hbox{$\refpos#1$}}

\begin{document}

\begin{titlepage}

\begin{flushright}
 DAMTP/R-96/9 \\
 gr-qc/9603050
\end{flushright}

\mbox{ } \\
\mbox{ } \\
\mbox{ } \\

\LARGE
\begin{center}
 {\bf The Gravitational Hamiltonian in the Presence of Non-Orthogonal 
       Boundaries}
\end{center}

\mbox{ } \\
\mbox{ } \\
\normalsize

\begin{center}
 S. W. Hawking\footnote{S.W.Hawking@damtp.cam.ac.uk} and 
 C. J. Hunter\footnote{C.J.Hunter@damtp.cam.ac.uk} \\ 
 \mbox{ }\\
 {\it Department of Applied Mathematics and Theoretical Physics} \\
 {\it Silver Street, Cambridge CB3 9EW} \\ 
 \mbox{ } \\
 February 28, 1996
\end{center}

\mbox{ } \\
\mbox{ } \\

\begin{center}
 {\bf Abstract}
\end{center}
  This paper generalizes earlier work 
  on Hamiltonian boundary terms by omitting 
  the requirement that the spacelike hypersurfaces $\Sigma_t$ intersect
  the timelike boundary $\cal B$ orthogonally.  The expressions for the 
  action and Hamiltonian are calculated and the required subtraction of a 
  background contribution is discussed.  
  The new features of a Hamiltonian 
  formulation with non-orthogonal boundaries are then 
  illustrated in two examples.

\begin{center}
\end{center}

\end{titlepage}

\section{Introduction}
 \label{sec:introduction} 
  There has recently been a renewed interest, \cite{Brown93a,Hawking95a}, 
  in the 
  Hamiltonian formulation of general relativity, primarily motivated by the 
  desire to extend its range of applicability and utility to include more 
  general boundary conditions, such as those which arise in black hole pair 
  creation \cite{Hawking95b}.  To this date, however, it has been generally
  assumed, given the standard 3+1 decomposition of spacetime, that the normal 
  to the spacelike hypersurfaces is orthogonal to the normal of the boundary
  of the spacetime (although the effect of non-orthogonal boundaries 
  has been considered, for example in 
  \cite{Hawking79}-\cite{Lau95}, where the action is
 modified in order to account for joints or corners where the boundaries may
  be nonorthogonal).  
  The purpose of this paper is to derive the gravitational
  action and Hamiltonian, including all the terms which arise from the 
  non-orthogonality of the boundaries, and illustrate the utility of such a 
 derivation by calculating two examples. 

  Let $({\cal M},g)$ be a sufficiently well-behaved four-dimensional spacetime, 
  admitting a scalar time function $t(x^\mu )$, from $\cal M$ onto $[0,1]$,
  which foliates
  $\cal M$ into a family of spacelike hypersurfaces, $\{ \Sigma_t\}$,
  of constant $t$.
  The boundary of $\cal M$ consists of the initial and final spacelike 
  hypersurfaces $\Sigma_0$ and $\Sigma_1$, as well as a timelike boundary
  $\cal B$, hereafter called the three-boundary.  
  For each $\Sigma_t$, we can define a two-surface, $B_t=\Sigma_t\cap \cal B$, 
  which bounds the hypersurface.  The family of two-surfaces $\{ B_t\}$ then
  foliates the three-boundary $\cal B$.  
  The spacetime and its submanifolds are shown
  in figure 1.  The tensors defined on the surfaces 
  are given in table 1, and are an amalgamation of the notation
  adopted in \cite{Brown93a} and \cite{Hawking95a}.  Greek letters are used
  for indices on $\cal M$, while middle roman letters ($i \cdots p$) are used 
  for indices
  on $\Sigma_t$, middle roman letters with a circumflex for indices on $\cal B$,
  and early roman letters ($a \cdots d$) are used for indices on $B_t$.  
  Tensors on any of the 
  submanifolds can also be considered as tensors in $\cal M$, by the obvious
  embedding, and in this context are denoted by greek indices.  However, care
  must be taken with raising and lowering indices.
  An index is considered to be lowered or
  raised by the metric corresponding to the type of index used.  
 \begin{figure}[htb]
  \hspace{1.8cm}
   \vbox{
    \beginlabels\refpos 139 742 {}
                \put 265 418 {\Sigma_0}
                \put 265 568 {\Sigma_t}
                \put 379 614 {n^\mu}
                \put 417 588 {u^\mu}
                \put 284 709 {\Sigma_1}
                \put 127 473 {B_t}
                \put 459 683 {\cal B}
    \endlabels
    \epsfxsize=.75\textwidth
    \epsfbox{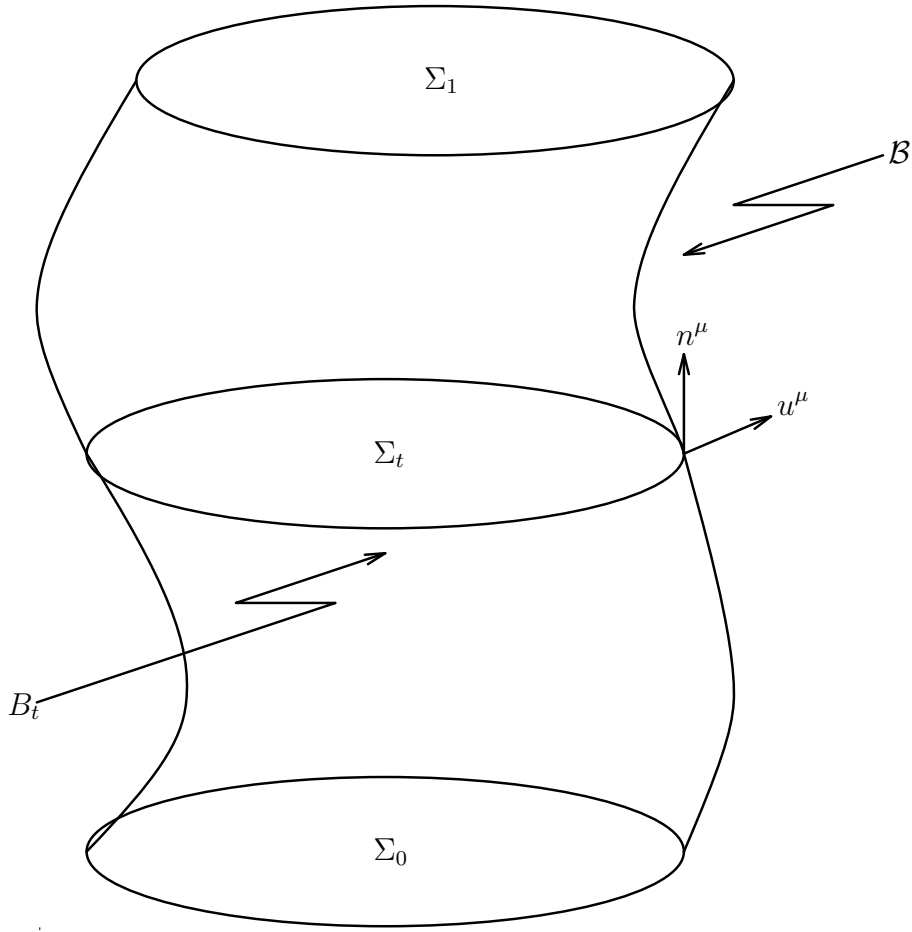}
    }
   \caption{The spacetime manifold $\cal M$ and its submanifolds are shown, 
   with one spacelike dimension suppressed.  The unit normals $n^\mu$, and
   $u^\mu$ to the spacelike hypersurface $\Sigma_t$ and the timelike 
   boundary $\cal B$ are shown at points on the two-surface $B_t$.}
   \label{fig:manifold}
  \end{figure}
 \begin{table}[htb]
 \begin{center}
  \begin{tabular}{lcccccc} \hline \hline
   & & Covariant & Unit & Intrinsic & Extrinsic & \\
   & Metric & derivative & normal & curvature & curvature & Momentum \\ \hline
   Spacetime $\cal M$ & $g_{\mu\nu}$ & $\nabla_\mu$ & & 
       $R_{\mu\nu\rho\sigma}$ & & \\
   & & & & & & \\
   Hypersurfaces $\Sigma_t$ & & & & & & \\
   \mbox{    } embedded in $\cal M$ & $h_{ij}$ & $D_i$ & $n_\mu$ & 
       ${\cal R}_{ijkl}$ & $K_{ij}$ & $P^{ij}$ \\
   & & & & & & \\
   Three-boundary $\cal B$ & & & & & & \\
   \mbox{    } embedded in $\cal M$ & $\gamma_{\hat{i}\hat{j}}$ & 
      ${\cal D}_{\hat{i}}$ & $u_\mu$ & 
        & $\Theta_{\hat{i}\hat{j}}$ & \\
   & & & & & & \\
   Two-boundaries $B_t$ & & & & & & \\
   \mbox{    } embedded in $\Sigma$ & $\sigma_{ab}$ &  & $r_i$ & 
       & $k_{ab}$ & \\ \hline \hline 
  \end{tabular}
  \caption{The naming conventions for the tensors on the spacetime $\cal M$, 
           and the hypersurfaces and surfaces embedded therein.}
  \label{tab:defn}
 \end{center}   
 \end{table}

  We will work in a coordinate system adapted to the time function and the
  three-boundary, that is, 
  the first coordinate is $t$, and ${\cal B}$ is a surface of constant $x^1$.  
  This allows us to write the metric $g_{\mu\nu}$ in the usual ADM 
  \cite{Arnowitt62a} decomposition, 
  \begin{equation}
   ds^2 = -N^2 dt^2 + h_{ij}(dx^i + N^i dt)(dx^j + N^j dt),
  \end{equation}
  where $N$ is the lapse function (taken to be the positive square root), and 
  $N^i$, which lies in the tangent plane of $\Sigma_t$, is the 
  shift vector. 
  We can define a timelike vector 
  field $t^\mu$, which we interpret as connecting corresponding points on 
  adjacent hypersurfaces,  
  \begin{equation}  
   \label{eqn:tmu}
   t^\mu = Nn^\mu + N^\mu,
  \end{equation}
  where $n^\mu$ is the unit normal to the spacelike hypersurfaces.
  Thus, given a point on a hypersurface, its evolution normal to the 
  hypersurface is governed by the lapse function,
  while the shift vector dictates its evolution tangent to the hypersurface.
  The orientations of the unit normals $n^\mu$ and $u^\mu$ are fixed by 
  requiring
  that they be future pointing and outward pointing, respectively. 
  The intersection of the two hypersurfaces is not required to 
  be orthogonal, since this would be an unnecessary restriction on the 
  spacetime.   The non-orthogonality is measured by using the variable  
  \begin{equation}
   \eta = n_\mu u^\mu,
  \end{equation} 
  which clearly vanishes in the orthogonal case.

  If we want $\cal B$ to be mapped into itself by 
  $t^\mu$, then clearly $t^\mu$ must lie in the three-boundary, and hence it 
  must be orthogonal to the normal $u^\mu$.  For this to occur, it is 
  necessary that the lapse function and shift vector satisfy  
  \begin{equation}
   \label{eqn:t_boundary}
   \eta = - \frac{u_\mu N^\mu}{N}. 
  \end{equation}
  But, more generally, 
  it is not necessary for $t^\mu$ to lie in the boundary, and indeed it is
  impossible if you are considering a spatial translation.

  In section \ref{sec:action} of the paper, the action and Hamiltonian are 
  calculated for metrics of Lorentzian signature.  In 
  order to obtain a finite answer for a non-compact spacetime it is necessary
  to consider the action and Hamiltonian relative to a background spacetime
  \cite{Gibbons77}.  This process is outlined in section \ref{sec:background}.  
  In section \ref{sec:examples}, two examples are presented in order to 
  illustrate some of the
  properties of the Hamiltonians obtained from spacetimes with non-orthogonal 
  boundaries. 
  The appendices contain various kinematical equations, some relations 
  based on the particular 
  coordinate system adopted in this paper, and a summary of the corresponding
  Euclidean results.

\section{Calculation of the Action and Hamiltonian}
 \label{sec:action}
 The standard Hilbert action for Lorentzian general relativity is  
 \begin{eqnarray}
  \label{eqn:action}
  I[{\cal M}, g] & = & \frac{1}{2\kappa} \int_{\cal M} d^4x\,\sqrt{-g}R
      + \frac{1}{\kappa} \int_{\Sigma_0}^{\Sigma_1} d^3x\,\sqrt{h}K 
      + \frac{1}{\kappa} \int_{\cal B} d^3x\,\sqrt{-\gamma}\Theta \nonumber \\
 & &  + \frac{1}{\kappa} \int_{B_0}^{B_1} d^2x\, \sqrt{\sigma} {\rm sinh}^{-1}
        \eta, 
 \end{eqnarray}
 where $\kappa$ is $8\pi G$, the integral between $\Sigma_0$ and $\Sigma_1$
 is notation for the
 integral over the final hypersurface, $\Sigma_1$, minus the integral over the
 initial hypersurface, $\Sigma_0$, and similarly for the integral over the 
 initial and final two-surfaces, $B_0$ and $B_1$,  
(which bound the initial and final hypersurfaces $\Sigma_0$ and $\Sigma_1$).
 The final term, referred to as the corner term, is 
 necessary in order to ensure that the variation of the action with respect to
 the intersection angle $\eta$ vanishes, \cite{Hayward93},  
 and is the only effect of 
 non-orthogonal boundaries which has been considered  previously.
 In order to obtain the Hamiltonian, we need to factor the integrals into 
 an integral over time $t$, and integrals over spacelike surfaces. 
 We would also like to separate out the terms in the volume integral 
 which are pure divergence, and
 convert them to boundary integrals.
 Let $I_{\cal M}$, $I_{\cal B}$, $I_\Sigma$, and $I_B$ be the integrals over 
 the corresponding manifolds in equation (\ref{eqn:action}).

 Using
 equation (\ref{eqn:curvature}), we can substitute, for the curvature scalar 
 $R$, terms which lie in the hypersurface $\Sigma_t$.  Thus, the volume 
 integral becomes   
 \begin{equation}
  I_{\cal M} = \frac{1}{2\kappa} \int_{\cal M} d^4x\,\sqrt{-g}[{\cal R} + 
	K_{\mu\nu}K^{\mu\nu} - K^2 + 2\nabla_\mu(n^\mu K - a^\mu )],
 \end{equation}
 where $a^\mu = n^\nu \nabla_\nu n^\mu$ is the acceleration of the unit normal
 $n^\mu$.
 We can
 convert the final term in the volume integral 
 to a surface integral over the boundary of $\cal M$.  Thus, $I_{\cal M}$ 
 simplifies to 
 \begin{equation}
  I_{\cal M}' = \frac{1}{2\kappa} \int_{\cal M} d^4x\,\sqrt{-g}[{\cal R} + 
	K_{\mu\nu}K^{\mu\nu} - K^2].
 \end{equation}
 The integral over the three-boundary becomes
 \begin{equation}
  I_{\cal B}' = \frac{1}{\kappa} \int_{\cal B} d^3x\,\sqrt{-\gamma}[\Theta 
       + u_\mu(n^\mu K - a^\mu)],
 \end{equation}
 while the integral over the initial and final hypersurfaces becomes
 \begin{equation}
  I_\Sigma '=  
       \frac{1}{\kappa} \int_{\Sigma_0}^{\Sigma_1} d^3x\,\sqrt{h}[K 
	  + n_\mu(n^\mu K - a^\mu)].
 \end{equation}
 But, since $n_\mu a^\mu$ vanishes, and the $K$ terms cancel, we see that 
 $I_\Sigma '$ disappears.  
 Hence, the action reduces to $I_{\cal M}' + I_{\cal B}' + I_B$,
 \begin{eqnarray}
  I & = & \frac{1}{2\kappa} \int dt \int_{\Sigma_t} d^3x\,N\sqrt{h}[{\cal R} + 
	K_{\mu\nu}K^{\mu\nu} - K^2] 
      + \frac{1}{\kappa} \int_{\cal B} d^3x\,\sqrt{-\gamma}[\Theta 
       + \eta K - u_\mu a^\mu] \nonumber \\
  & &     + \int_{B_0}^{B_1} d^3x\, \sqrt{\sigma} {\rm sinh}^{-1} \eta.
 \end{eqnarray}
 We now want to maneuver the action into canonical form.  
 Using equation (\ref{eqn:integrand}), we can write 
 $I_{\cal M}'$ in terms of canonical variables, 
 \begin{equation}
  I_{\cal M}' = \int dt \int_{\Sigma_{t}} 
                      d^3x\,[P^{\mu\nu}\dot{h}_{\mu\nu} - N{\cal H} 
	- 2P^{\mu\nu}D_\mu N_\nu], 
 \end{equation}
 where $\cal H$ is the energy constraint, which vanishes on any solution of the
 field equations.
 From equations (\ref{eqn:two_curvature}) and (\ref{eqn:gamma_sigma}) 
 we can write $I_{\cal B}'$ as 
 \begin{equation}
  I_{\cal B}' = \frac{1}{\kappa} \int dt \int_{B_t} 
       d^2x\,N\sqrt{\sigma} [k
      - \lambda^2 v^\mu \nabla_\mu \eta  ], 
 \end{equation}
 where $v^\mu = \lambda(n^\mu - \eta u^\mu)$ is the normalized 
 projection of $n^\mu$ onto the three-boundary, and  
 $\lambda = (1+\eta^2)^{-\frac{1}{2}}$ is the normalization constant for
 the unit vectors $r^\mu$ and $v^\mu$. 
 We can write the nonorthogonal part of the boundary integral as the sum of 
 a total derivative and a second term,
 \begin{equation}
 \label{eqn:total_tilt}
  I_{\cal B}' = \frac{1}{\kappa} \int dt \int_{B_t} d^2x\,[N\sqrt{\sigma} k
      - \sqrt{-\gamma} (\nabla_\mu (v^\mu {\rm sinh}^{-1} \eta ) + 
     {\rm sinh}^{-1} \eta \nabla_\mu v^\mu ) ]. 
 \end{equation}
 If we convert the total derivative to a boundary integral over $B_0$ and $B_1$,
 then it will cancel $I_B$, and the action will simply be the sum of the 
 integrals over $\cal M$ and $\cal B$, where the three-boundary integral is now
 \begin{equation}
  I_{\cal B}'' = \frac{1}{\kappa} \int dt \int_{B_t} d^2x\,N\sqrt{\sigma} ( k
      + \lambda {\rm sinh}^{-1} \eta \nabla_\mu v^\mu  ).
 \end{equation}

 Finally, to obtain the action in canonical form, we need to remove from 
 $I_{\cal M}'$  the term 
 involving the derivative of the shift vector. 
  Using equation (\ref{eqn:mom_constraint}), we obtain 
 \begin{equation}
  I_{\cal M}' = \int dt \int_{\Sigma_t} 
              d^3x\,[P^{\mu\nu}\dot{h}_{\mu\nu} - N{\cal H} 
	- N^\mu{\cal H}_\mu - 2D_\mu(P^{\mu\nu}N_\nu)], 
 \end{equation}
 where $\cal H_\mu$ is the momentum constraint, which also 
 vanishes on any solution 
 of the field equations.
 Converting the final term into a surface integral over $B_t$, the
 volume contribution reduces to 
 \begin{equation}
  I_{\cal M}'' = \int dt \int_{\Sigma_t} 
     d^3x\,[P^{\mu\nu}\dot{h}_{\mu\nu} - N{\cal H} - N_\mu{\cal H}^\mu],
 \end{equation}
 while the boundary term becomes 
 \begin{equation}
  I_{\cal B}''' = 
       \frac{1}{\kappa} \int dt \int_{B_t} d^3x\,N\sqrt{\sigma} 
       [k  
      + \lambda {\rm sinh}^{-1} \eta \nabla_\mu v^\mu  ], 
      -2 \int dt \int_{B_t} d^2x\,r_\mu P_{\sigma}^{\mu\nu}N_\nu , 
 \end{equation}
 where $P_{\sigma}^{\mu\nu}$ is the tensor density $P^{\mu\nu}$ with the 
 correct area element for $B_t$, as given by equation (\ref{eqn:two_mom}). 
 Thus, by factoring out the time integral,
 we can express the action in canonical form,
 \begin{equation}
  I = \int dt \left\{ \int_{\Sigma_t} d^3x\,P^{\mu\nu}\dot{h}_{\mu\nu} - 
       (H_c + H_k + H_t + H_m)
      \right\}, 
 \end{equation}
 where the Hamiltonian, 
 \begin{equation}
  \label{eqn:hamiltonian}
  H[{\cal M}, g] = H_c + H_k + H_t + H_m, 
 \end{equation}
 is a sum of four distinct terms:
 \begin{enumerate}
  \item a constraint term, 
   \begin{equation}
    H_c = \int_{\Sigma_t} d^3x\,[ N{\cal H} 
	+ N_\mu{\cal H}^\mu ],
   \end{equation}
  \item a curvature term, 
   \begin{equation}
    \label{eqn:Hcurvature}
    H_k =  
      - \frac{1}{\kappa} 
      \int_{B_t} d^2x\,N\sqrt{\sigma}k,
   \end{equation}
  \item a tilting term, 
   \begin{equation}
    H_t = 
       - \frac{1}{\kappa} \int_{B_t} d^2x\,N\lambda\sqrt{\sigma}
        {\rm sinh}^{-1} \eta \nabla_\mu v^\mu , 
   \end{equation}
  \item and a momentum term, 
   \begin{equation}
    \label{eqn:Hmomentum}
    H_m = 
       2 \int_{B_t} d^2x\,r_\mu P_{\sigma}^{\mu\nu}N_\nu . 
   \end{equation}
 \end{enumerate}

 In the case of orthogonal boundaries, the curvature term, $H_k$, is usually 
 taken to give the mass of the system, while the momentum term, $H_m$, gives
 the linear or angular momentum.  However, it will be shown in section 
 \ref{sec:examples} that the two contributions 
 get mixed up in the non-orthogonal
 case.  

 We also see that the only place in which the non-orthogonality appears
 explicitly is in the tilting term, $H_t$.  One can see that the action must
 depend on the way one chooses the angle between $\Sigma_t$ and the 
 three-boundary to vary with time.  If the resulting two-surfaces, $B_t$, are 
 independent of time, as they would be on an inner horizon, then the 
 tilting term vanishes, 
 since the total non-orthogonal contribution to the integral over the 
 three-boundary is simply the 
 negative time derivative of the corner term, and hence the two terms will 
 cancel. 
 If the two-surface is time-dependent, as it can be on the boundary at infinity,
 then $H_t$ will in general be nonzero.  However, this term will cancel when 
 we subtract the action of the background, as detailed in the next section.
 Since the action must be independent of the way in which we choose our 
 boundaries to intersect, we see that a Hamiltonian treatment of the action 
 will not work unless we include a background subtraction to remove the tilting
 term.
 
 \section{Background Spacetime}
  \label{sec:background}
  If the time surfaces $\Sigma_t$ are non-compact, the action is calculated
  by evaluating the action on a compact region, and then letting the boundary
  tend to infinity.  This is problematic, since the Hamiltonian will 
  generally diverge as the boundary is taken to infinity.  However, it makes
  sense to define the physical Hamiltonian, $H^P$, to be the difference 
  between the 
  Hamiltonian for the space under consideration, and the Hamiltonian for some
  background solution of the field equations, which can be regarded as a ground 
  state for solutions with that asymptotic behaviour.  Quantities defined on the
  background spacetime are indicated by a tilde.  A minimum requirement for
  a solution to be regarded as a ground state would seem to be that it had a
  timelike Killing vector, but one might also ask that it was completely 
  homogeneous and had three linearly independent Killing vectors as well.  
  The existence of the timelike Killing vector is necessary for a suitable
  definition of energy.
  The usual background is Minkowski space, but one can consider other 
  backgrounds, such as anti-de Sitter space \cite{Hawking73}, the 
  Robinson-Bertotti metric \cite{Robinson59}, or
  the Melvin universe \cite{Melvin64}.  The last is not homogeneous, but has 
  been used as the background metric for the pair creation of charged black 
  holes \cite{Hawking95b}.  Choosing an appropriate background for 
  Kaluza-Klein spacetimes presents additional difficulties 
 \cite{Bombelli87,Deser89}, due to the presence of the compactified dimensions.
 These spacetimes will not be considered here, but will be addressed in 
 forthcoming papers \cite{Hawking96,Chamblin96}.  

  The induced metrics on the three-boundary should agree in the two solutions,
  but in some cases they will agree only asymptotically, in the limit that the 
  three-boundary goes to infinity.\footnote{If the spacetime has a horizon, then
  it may be possible to analytically continue the solution through the horizon 
  to obtain a second asymptotic region of space.  In this case, we want to take
  $\Sigma_t$ to have an inner boundary on the horizon, rather than two 
  asymptotic regions.  Alternatively, we could use a star, or similar 
  physical object, to eliminate the
  extra asymptotic region.}  It is a delicate matter to choose the rate 
  at which the metrics on the boundaries approach each other, and it will 
  depend on the asymptotic behaviour of the lapse function and shift vector
  under consideration.  We shall assume that the necessary fall-off conditions
  are satisfied, and shall therefore take the induced metrics on the boundary
  to agree.

  Since the background metric is taken to be a solution of the field equations,
  the physical Hamiltonian will only have a constraint contribution from the
  actual spacetime (and only if it is not a solution),
  \begin{equation}
    H^P_c = \int_{\Sigma_t} d^3x\,[ N{\cal H} + N_\mu{\cal H}^\mu ].
   \end{equation}
  Because the induced metrics of the actual spacetime and its background agree 
  on the three-boundary, they will have the same volume element, 
  $N\lambda\sqrt{\sigma}$.  Furthermore, we can always take the angle between
  $\Sigma_t$ and ${\cal B}$ to be the same for the two spacetimes, hence 
  the respective values of $\lambda$ will also be the same.  Thus, the 
  curvature term is simply  
   \begin{equation}
    \label{eqn:physCurv}
    H^P_k =  - \frac{1}{\kappa} \int_{B_t} d^2x\,
     N\sqrt{\sigma}\left( k - \tilde{k} \right).  
   \end{equation}
  If the background slices are chosen such that the conjugate momentum 
  vanishes, then the momentum contribution to the Hamiltonian will simply
  be from the spacetime under consideration.  However, the constraint of 
  matching $\eta$ in both spacetimes implies that we cannot always ensure that
  the momentum density vanishes in the background.  Thus,  
   \begin{equation}
    H^P_m = 2  \int_{B_t} d^2x\, \left( r_\mu P_{\sigma}^{\mu\nu}N_\nu - 
         \tilde{r}_\mu \tilde{P}_{\sigma}^{\mu\nu} \tilde{N}_\nu \right).
   \end{equation}
  Since $\eta$ is the same for both spacetimes, and the vector $v^\mu$ lies 
  entirely in $\cal B$, the tilting terms will be equal for both 
  spacetimes, and hence will cancel each other when subtracted.  Thus, there
  will be no contribution to the physical Hamiltonian from the tilting term, 
   \begin{equation}
    H^P_t  =  0.
   \end{equation}
 The vanishing of the tilting term is necessary for the Hamiltonian 
 formulation to be well-defined, because otherwise we have an unacceptable 
 dependence of the action on the way in which the boundaries intersect.
  Note that on an inner boundary, such as a horizon, there may not be a  
  subtraction, since the background may not have a horizon, but, as noted at the
 end of section \ref{sec:action}, the tilting term will vanish anyway, since 
 the horizon two-surfaces are time-independent.

  For a Killing vector, $\xi^\mu$, of the background spacetime, we can obtain a
  conserved charge on the spacelike hypersurfaces by decomposing $\xi^\mu$ in
  terms of a lapse function and shift vector, and calculating the corresponding
  Hamiltonian.  Thus, assuming that $t^\mu$ is asymptotically equal to the 
  time translation Killing vector of the background spacetime, the energy, $E$,
  which is the conserved charge associated with time translation,
  is simply the value of the physical Hamiltonian, $H^P$.

 \section{Examples}
  \label{sec:examples}
  \subsection{Schwarzschild Spacetime with Flat Spacelike Slices}

   We first want to consider a simple example which has a non-orthogonal 
   intersection of boundaries, but for which the tilting term $H_t$ 
   vanishes, and where there is no non-trivial spatial linear momentum.  
   We begin
   by computing the terms in the physical Hamiltonian, first for the 
   Schwarzschild spacetime, and then for the background Minkowski space.  Two 
   different background coordinate systems are used--a minimally matched system
   in which the induced metrics agree, but the values of $\eta$ differ, and a
   correctly matched system in which both the induced metric and $\eta$ agree.
   Once the terms 
   have been computed, we can obtain the physical Hamiltonian.  
   The calculations of this example and the
   next one both made extensive use
   of the GRTensorII package for Maple \cite{Musgrave94}.   

   If we consider the Schwarzschild solution in standard static coordinates, 
 \begin{equation}
    ds^2 = -(1-\frac{2M}{r})dt^2 + \frac{1}{1-\frac{2M}{r}} dr^2 + r^2d\Omega,
 \end{equation}
   and then define a new time coordinate $t'$ by
 \begin{equation}
    dt = dt' - \frac{\sqrt{\frac{2M}{r}}}{1-\frac{2M}{r}}dr,
 \end{equation}
   then the Schwarzschild line element becomes
   \begin{eqnarray}
    ds^2 & = & -{dt'}^2 + (dr + \sqrt{\frac{2M}{r}}dt')^2 + r^2d\Omega \\
         & = & -(1-\frac{2M}{r}){dt'}^2 + 2\sqrt{\frac{2M}{r}}dr\,dt' + 
                dr^2 + r^2d\Omega,
   \end{eqnarray}
   which is the Painleve and Gullstrand \cite{Painleve21} line element, most 
   recently investigated in \cite{Kraus94a}.  This coordinate system is 
   of interest primarily because it has flat spacelike slices and the 
   intersection of a hypersurface of constant time and one of constant 
   radius is non-orthogonal. 
   The lapse function and the radial component of the shift vector are 
   non-vanishing, 
 \begin{equation}
    N = 1, \hspace{.5cm} {\rm and} \hspace{.5cm} N^{r} = \sqrt{\frac{2M}{r}}.
 \end{equation}
 As stated in the introduction, we want to consider spacelike hypersurfaces 
 $\Sigma_{t'}$ of constant time $t'$, and a three-boundary $\cal B$ of constant
 radius $r$.  In order to calculate the
 physical Hamiltonian, we need to compute the Hamiltonian for a fixed
 three-boundary which we then let tend to infinity. 
 Hence, take the fixed three-boundary to be the hypersurface
 of radius $R$, denoted ${\cal B}^R$.  
 Since the calculation is independent of the choice of
 spacelike hypersurface, we do not need to fix $\Sigma_{t'}$.  
 We find that $t^\mu u_\mu$ vanishes, and hence $t^\mu$ lies in the 
 three-boundary, so that  
 ${\cal B}^R$ is evolved into itself by $t^\mu$. 
 The intersection of $\Sigma_{t'}$ and ${\cal B}^R$ is not 
 orthogonal, but is characterized by the variables 
 \begin{equation}
    \eta = -\sqrt{\frac{2M}{R-2M}}, \hspace{.5cm} {\rm and} \hspace{.5cm}
    \lambda = \sqrt{1 - \frac{2M}{R}}.
 \end{equation}
 The two-surface $B^R_{t'} = \Sigma_{t'} \cap {\cal B}^R$ is a 
 sphere of constant $t'$ and $r$.

 As noted above, the induced metric on $\Sigma_{t'}$ is completely 
 flat, and hence the induced metric on $B^R_{t'}$ is simply 
 the standard two-sphere metric.  Thus, these metrics have identical volume
 elements, while the volume element of the three-boundary ${\cal B}^R$ 
 contains a contribution from the non-trivial value of $\lambda$,
   \begin{equation}
    \sqrt{h} = R^2\sin\theta = \sqrt{\sigma}, \hspace{ .5cm} {\rm and}
    \hspace{.5cm}
    N\lambda\sqrt{\sigma} = \sqrt{1-\frac{2M}{R}}R^2\sin\theta.
   \end{equation}

 We now would like to calculate the terms which contribute to the Hamiltonian.
 The trace of the extrinsic curvature of the two-surface, $B^R_t$, is
 \begin{equation}
  \label{eqn:ex1fork}
  k = \frac{2}{R}.
 \end{equation}
 The two-surface $B^R_{t'}$ is independent of time, 
 so $\nabla_\mu v^\mu$ vanishes,
 and hence there will be no tilting contribution to the Hamiltonian. 
 If we calculate the momentum tensor density on $B^R_{t'}$, 
 $P_\sigma^{\mu\nu}$, and contract it with the unit normal, $r^\mu$, and the
 shift vector, $N^\mu$, we obtain 
 \begin{equation}
  \label{eqn:ex1form}
    r_\mu P_\sigma^{\mu\nu} N_\nu = \frac{2M}{\kappa}\sin\theta,
 \end{equation}
 which is independent of the fixed radius $R$.

 The natural choice for the background spacetime is Minkowski space, since it
 has the same asymptotic behaviour as the Schwarzschild solution.
 We first consider a coordinate system in which the induced metric on the 
 three-boundary of radius $R$ agrees with the Schwarzschild case, but the 
 intersection angle, characterized by $\eta$, is different.  The correct
 background coordinate system, with matched $\eta$ values, is then used.  The
 Hamiltonian is identical in both cases, but the contributions from each of the
 terms depends on the coordinate system chosen.
 
 Consider the static coordinate system with a scaled time coordinate,
 \begin{equation}
  \label{eqn:MinkBack}
  ds^2 = -(1-\frac{2M}{R})dt^2 + dr^2 + r^2d\Omega^2.
 \end{equation}
 As in the Schwarzschild case, we consider a fixed three-boundary of radius 
 $R$, $\tilde{{\cal B}}^R$.
 The induced metric on $\tilde{{\cal B}}^R$ is then the same as the induced 
 metric on the three-boundary ${\cal B}^R$, and hence we can equate the two
 boundaries.  
 The lapse function and shift vector are not the
 same as in the Schwarzschild case, but instead 
 \begin{equation}
     \tilde{N} = \sqrt{1-\frac{2M}{R}},
 \end{equation}
 while the shift vector vanishes.  The disagreement in the lapse is due to the
 disagreement between the boundary intersection value $\eta$ in the two 
 solutions.
 Since the boundaries in the background coordinate system were chosen to be
 orthogonal, $\tilde{\eta}$ vanishes, and 
 $\tilde{\lambda}$ reduces to one.  

 We now want to calculate the terms contributing to the Hamiltonian.
 The trace of the extrinsic curvature of the two-boundary is
 \begin{equation}
  \label{eqn:ex1backk}
  \tilde{k} = \frac{2}{R}.
 \end{equation}
 As in the Schwazrschild case, the two-surface $B^R_{t'}$ 
 is time-independent, and
 hence $\nabla_\mu v^\mu$ is zero.
 The momentum density $\tilde{P}^{\mu\nu}$ also vanishes, so that
 \begin{equation}
  \label{eqn:ex1backm}
  \tilde{r}_\mu \tilde{P}^{\mu\nu}_\sigma \tilde{N}_\nu = 0.
 \end{equation}

 We can now calculate the physical Hamiltonian.
 Since both the Schwarzschild and Minkowski spacetimes are solutions of 
 the field equations, the constraint term vanishes.
 The curvature contribution to the physical Hamiltonian 
 can be calculated by integrating equations (\ref{eqn:ex1fork}) and 
 (\ref{eqn:ex1backk}) over the two-surface $B^R_t$ (taking care to include the
 factors arising from the difference in $\lambda$ values between the two 
 solutions), 
 \begin{equation}
    H^P_k = -\frac{1}{\kappa} \lim_{R \rightarrow \infty} \int_0^{2\pi} d\phi 
               \int_{-1}^{1} d(\cos \theta)\, \sqrt{1-\frac{2M}{R}} R^2
               \left( \frac{2}{R}\frac{1}{\sqrt{1-\frac{2M}{R}}} - 
                      \frac{2}{R} \right).
 \end{equation}
 If we expand the square root in a Taylor series, then
 the infinite contributions cancel, and we are left with only a finite value,
 \begin{equation}
    H^P_k = -M. 
 \end{equation}
 The tilting contribution is zero because $\nabla_\mu v^\mu$ vanishes in
 both the Schwarzschild and Minkowski cases.
 The momentum contribution is due entirely to the Schwarzschild term, equation
 (\ref{eqn:ex1form}),  
 \begin{equation}
    H_m^P = 2 \int_0^{2\pi} d\phi 
          \int_{-1}^1d(\cos \theta )\frac{2M}{\kappa}
        = 2M.
 \end{equation}
 Thus, the Hamiltonian is
 \begin{equation}
  \label{eqn:mass}
  H^P = H^P_k + H^P_m = M, 
 \end{equation}
 as anticipated.  Note that there were contributions from both $H^P_k$ and 
 $H^P_m$,
 contrary to standard expectation.  However, as will be shown, this is due to 
 incorrect matching of the background.

 We would now like to consider a background coordinate system in which the 
 value of $\eta$ agrees with that of the Schwarzschild solution. 
 If we introduce a new time coordinate
 \begin{equation}
  dt = dt' - \frac{\sqrt{\frac{2M}{R}}}{\sqrt{1-\frac{2M}{R}}} dr,
 \end{equation}
 then the line element (\ref{eqn:MinkBack}) becomes
 \begin{equation}
  ds^2 = -(1 - \frac{2M}{R}) dt'^2 + 
         2\sqrt{\frac{2M}{R}}\sqrt{1- \frac{2M}{R}}dt'dr +
         (1- \frac{2M}{R})dr^2 + r^2 d\Omega^2.
 \end{equation}
 The lapse function and radial shift vector are non-vanishing,
 \begin{equation}
    \tilde{N} = 1, \hspace{.5cm} {\rm and} \hspace{.5cm} 
    \tilde{N^{r}} = \sqrt{\frac{2M}{R}}\sqrt{1 - \frac{2M}{R}},
 \end{equation}
 and the lapse agrees with the lapse of the Schwarzschild case.
 The induced metrics on a three-boundary of fixed radius $R$ still agree,
 but now the boundary intersection is non-orthogonal, since
 \begin{equation}
    \tilde{\eta} = -\sqrt{\frac{2M}{R-2M}},\hspace{.5cm}{\rm and}\hspace{.5cm}
    \tilde{\lambda} = \sqrt{1 - \frac{2M}{R}},
 \end{equation}
 which are equal to the Schwarzschild values.  If we calculate the trace of the 
 extrinsic curvature of the two-surface, $B^R_{t'}$, we obtain
 \begin{equation}
  \label{eqn:ex1bMinkk}
  \tilde{k} = \frac{1}{R\sqrt{1-\frac{2M}{R}}}.
 \end{equation} 
 The conjugate momentum is now non-trivial in this coordinate system, 
 \begin{equation}
  \label{eqn:ex1bMinkM}
  \tilde{r}_\mu (\tilde{P}_\sigma)^{\mu\nu} \tilde{N}_{\nu} 
     = \frac{2M}{\kappa}\sin\theta.
 \end{equation}
 We see that just as in the other coordinate system, 
 $\nabla_\mu v^\mu$ vanishes.

 We now want to calculate the physical Hamiltonian.  To calculate the 
 curvature contribution, we integrate equations (\ref{eqn:ex1fork}) and
 (\ref{eqn:ex1bMinkk}) over the two-surface (where now since the values of
 $\lambda$ agree, we can use equation (\ref{eqn:physCurv})),   
 \begin{equation}
  H^P_k = -\frac{1}{\kappa} \lim_{R \rightarrow \infty} \int_0^{2\pi} d\phi\, 
            \int_{-1}^{1} d(\cos \theta)\, R^2
            \left[ \frac{2}{R} - \frac{2}{R}\frac{1}{\sqrt{1 - \frac{2M}{R}}}
            \right].
 \end{equation}
 When we expand the square root in a Taylor series, we obtain
 \begin{equation}
  H^P_k = M.
 \end{equation}
 Since the two momentum terms, (\ref{eqn:ex1form}) and (\ref{eqn:ex1bMinkM})
 are identical, they will cancel when subtracted, and hence $H_m$ will vanish.
 As in the previous case, $\nabla_\mu v^\mu$ disappears in both the 
 Schwarzschild and Minkowski spacetimes, and hence there will be no tilting
 term contribution to the Hamiltonian.  
 Thus, the only contribution to the physical Hamiltonian is from the
 curvature term, 
 \begin{equation}
  H^P = H^P_k = M,
 \end{equation}
 as anticipated. 
 Note that
 this is an example where the constraint of matching $\eta$ between the
 spacetimes means that we
 cannot pick the background spacetime such that $\tilde{P}^{\mu\nu}$ vanishes. 
 This is contrary to the situation when only non-orthogonal boundaries are
 considered.
 It is obvious from the symmetry of the spacetime that the linear and angular 
 momenta vanish.

 As noted in section \ref{sec:background}, in order to avoid a second 
 asymptotic region we must consider an inner boundary on the horizon, at 
 $r=2M$.  There is no background contribution, since Minkowski 
 space has no horizon there, and hence the result will be the same for both 
 choices of background coordinates.  If we calculate the curvature contribution
 at the horizon, we obtain
 \[
  H^P_k = -2M,
 \]
 while the momentum contribution (which was noted to be independent of radius) 
 is
 \[
  H^P_m = 2M.
 \]
 Thus, we see that there is no net contribution from the horizon. 
 
 \subsection{Schwarzschild Spacetime with Tilted Spacelike Slices}

  In this example, we consider a more complicated slicing of the 
  Schwarzschild spacetime, corresponding to a Lorentz time boost in the $z$ 
  direction, which leads to a non-trivial tilting term.  In
  addition, we obtain a non-vanishing value for the spatial linear momentum.  
   
 If we take the Schwarzschild solution in static coordinates,
 and make the variable substitution,
 \begin{equation}
  t'  =  \frac{1}{\sqrt{1-v^2}}\left( t - vr\cos \theta \right), 
 \end{equation}
 then the line element becomes
 \begin{equation}
  ds^2 = - \left( 1 - \frac{2M}{r} \right) \left[ \sqrt{1-v^2}dt' + 
      v\cos \theta dr - vr\sin \theta d\theta \right]^2 +
           \frac{1}{1- \frac{2M}{r}}dr^2 
         + r^2d\Omega. 
 \end{equation}
 The lapse function is given by
 \begin{equation}
  N = \sqrt{\frac{r(r-2M)(1-v^2)}{r^2 - v^2(r-2M)(r-2M\cos\theta^2)}},
 \end{equation}
 while the shift vector has nonzero components in the $r$ and $\theta$ 
 directions,
 \begin{equation}
  N_r = -v\sqrt{1-v^2}\left[ 1 - \frac{2M}{r} \right] \cos\theta, 
   \hspace{.5cm} {\rm and} \hspace{.5cm}      
  N_\theta = v\sqrt{1-v^2}\left[ 1 - \frac{2M}{r} \right] r\sin\theta .
 \end{equation}
 As before, we take $\Sigma_{t'}$ to be the hypersurface of constant $t'$, 
 and we fix the three-boundary, ${\cal B}^R$,  to be the hypersurface of 
 constant radius $R$.  The intersection parameter $\eta$ is
 \begin{equation}
  \eta = \frac{v(r-2M)\cos\theta}{\sqrt{r^2 - v^2(r^2- 2Mr(1- \cos^2\theta) + 
       4M^2\cos^2\theta)}}.
 \end{equation}
 The two-surface $B^R_{t'}$ is simply a two-sphere of constant radius $R$ and
 time $t'$.  

 We now want to calculate the quantities which contribute to the physical 
 Hamiltonian.
 Expanding the relevant expressions in Taylor series, only terms which yield a
 non-vanishing value, in the limit as $R$ goes to infinity,
 are given.  To simplify the notation, we set $x = \cos\theta$. 
  The extrinsic curvature of the three-boundary, $B_{t'}^R$, is
 \begin{equation}
   \label{eqn:ex2kSC}
   k = \frac{(2-v^2[1-x^2])\sqrt{1-v^2}}{(1-v^2[1-x^2])^{3/2}}\frac{1}{R} - 
 \frac{2(1-v^2)-x^2v^4(1-x^2)}{\sqrt{1-v^2}(1-v^2[1-x^2])^{5/2}}\frac{M}{R^2}. 
 \end{equation}
  The derivative of $\eta$ in the direction of $v^\mu$ is
 \begin{equation}
   \label{eqn:ex2tilt}
 \nabla_\mu v^\mu = - \frac{vx(2-[1-\frac{2M}{r}][1-x^2]v^2)}{r^{3/2}(1 - 
   [1-\frac{2M}{r}][1-x^2]v^2)}.
 \end{equation}
  If we calculate the conjugate momentum, and contract it with $r^\mu$ and the
  shift vector, we obtain
  \begin{equation}
   \label{eqn:ex2SCm}
   r_\mu P_\sigma^{\mu\nu} N_\nu = -\frac{M}{2\kappa}\sin\theta 
   \frac{v^2(3x^2+1)}{\sqrt{1-v^2}}.
  \end{equation}

  We can now define the background spacetime, and calculate its contribution
  to the physical Hamiltonian.
  We need to find a coordinate system such that the induced metric on a
  boundary of constant radius $R$ agrees with the induced metric due to the
  Schwarzschild solution.  
  If we start with Minkowski space with scaled time, 
  given by equation (\ref{eqn:MinkBack}),
  and then make the coordinate transformation, 
 \begin{equation}
  t' = \frac{1}{\sqrt{1-v^2}}\left( t' - vR\frac{f(r)}{f(R)}\cos\theta \right),
 \end{equation}
 where 
 \begin{equation}
  f(r) = \frac{r}{2M}\left( 1 + \sqrt{1-\frac{2M}{r}}\right)^2
               e^{-2\sqrt{1-\frac{2M}{r}}},
 \end{equation}
 then the resulting coordinate system gives the same induced metric, value 
 of $\eta$, and lapse function on the three-boundary of constant radius $R$ as
 those found in the Schwarzschild metric. 
 If we calculate the trace of the extrinsic curvature of the two-surface, we
 find that
 \begin{equation}
  \tilde{k}=\frac{(2-v^2[1-x^2])\sqrt{1-v^2}}{(1-v^2[1-x^2])^{3/2}}\frac{1}{R}- 
         \frac{(2(1-v^2)-x^2(4-v^2[3-x^2]))v^2}
     {\sqrt{1-v^2}(1-v^2[1-x^2])^{5/2}}\frac{M}{R^2}. 
 \end{equation}
 Thus, if we integrate the difference between the Schwarschild value, given by 
 equation (\ref{eqn:ex2kSC}), and this background value, then the infinite 
 parts cancel, and we obtain the curvature contribution,
 \begin{equation}
  H^P_k = \frac{M}{\sqrt{1-v^2}}.
 \end{equation}
 The conjugate momentum, contracted with the unit normal $\tilde{r}^\mu$ and 
 the shift vector, is
 \begin{equation}
  \tilde{N}_\mu\tilde{P}_\sigma^{\mu\nu}\tilde{r}_\nu = 
        - \frac{M}{2\kappa}\sin\theta\frac{3x^2-1}{\sqrt{1-v^2}}.
 \end{equation}
 We see that if we integrate this over the two-surface it vanishes, and hence 
 the momentum contribution comes entirely from integrating Schwarzschild term,
 given by equation (\ref{eqn:ex2SCm}),
 \begin{equation}
  H^P_m = -\frac{Mv^2}{\sqrt{1-v^2}}.
 \end{equation}
 Finally, we note that, by construction, the tilting terms are equal 
 in the Schwarzschild and Minkowski systems, and hence cancel each other.  Thus, as expected, there is no
 tilting contribution to the Hamiltonian.  Hence, if we add the curvature and 
 momentum values, we obtain the value of the physical Hamiltonian, 
 \begin{equation}
  H^P = H^P_k + H^P_m = M\sqrt{1-v^2}.
 \end{equation}
 From its static value, the Hamiltonian has been decreased by the inverse of 
 the boost factor.  As will be shown below, this decrease in energy is 
 accounted for by a non-zero value of the linear momentum in the $z$ direction.
 Note that unlike the case when orthogonal boundaries are used, the physical 
 Hamiltonian now contains a non-trivial contribution from the momentum term.
 On the horizon the shift vector, the tilting term and the curvature term 
vanish, so it provides no contribution to the Hamiltonian.  

  We now want to consider the conserved charges arising from the Killing fields
  of the background spacetime. 
  The asymptotic value of the 
  background spacetime, in Cartesian coordinates, is
  \begin{equation}
   ds^2 = -(1-v^2)dt'^2  - 2v\sqrt{1-v^2}dt'dz + dx^2 + dy^2 + (1-v^2)dz^2.
  \end{equation} 
  This is related to the standard static background by the transformation
  \begin{equation}
   \label{eqn:transform}
   t' = \frac{1}{\sqrt{1-v^2}}(t - vz).
  \end{equation} 
  We can associate four Killing fields, $\tilde{t}^\mu$, and $\tilde{x}^\mu_i$
  with the translation invariance of each of the coordinates.  
  Since $\tilde{t}^\mu$ is
  equal to the time evolution vector $t^\mu$, the conserved charge 
  associated with time translation, ${\cal P}_t$, is simply the value of the
  physical Hamiltonian, $M\sqrt{1-v^2}$.
  If we consider the spatial Killing vectors $\tilde{x}_i^\mu$, 
  we find that by symmetry, the linear momenta in the $x$ and $y$ directions
  vanishes, but due to the transformation (\ref{eqn:transform}), there is 
  a nonzero value of the $z$ momentum.  The $z$ momentum will contain only a
  contribution from the momentum integral.  If we contract the conjugate 
 momentum with the unit normal and the Killing vector, we obtain
 \begin{equation}
  \tilde{z}_\mu P_\sigma^{\mu\nu} r_\nu = Mv(3x^2+1).
 \end{equation}
 The corresponding background value is
 \begin{equation}
  \tilde{z}_\mu \tilde{P}_\sigma^{\mu\nu} \tilde{r}_\nu = Mv(3x^2-1).
 \end{equation}
 Thus, the background contribution will integrate to zero, while the 
 Schwarzschild term yields 
 \begin{equation}
   {\cal P}_z =  \int d^2x\,\tilde{z}_\mu P_\sigma^{\mu\nu} r_\nu = Mv.
 \end{equation}
 The resulting energy momentum vector is
  \begin{equation}
   {\cal P}_\mu = \left( M\sqrt{1-v^2}, 0, 0, Mv \right).
  \end{equation}
 If we calculate the norm of the vector, with respect to the asymptotic
 background Minkowski space, we obtain 
  \begin{equation}
   {\cal P}_\mu {\cal P}^\mu = -M^2,
  \end{equation}
 indicating that it has transformed correctly as a four-vector.

 \section{Acknowledgements}
  C.J.H. acknowledges the financial support of the Association of Commonwealth
  Universities and the Natural Sciences and Engineering Research Council of
  Canada.

\appendix 

 \section{Kinematics}

  In this appendix, various standard kinematical formulae are presented, and 
  some relations used in the paper are derived.
 
 \subsection{The hypersurfaces $\Sigma_t$}

  The basic quantities which are induced on the hypersurface $\Sigma_t$ by
  the metric $g_{\mu\nu}$ and the unit normal $n^\mu$ are the first and second 
  fundamental forms, 
  generally called the induced metric and the extrinsic curvature,
  \begin{eqnarray}
   h_{\mu\nu} & = & g_{\mu\nu} + n_\mu n_\nu, \\
   K_{\mu\nu} & = & {h_{\mu}}^\alpha \nabla_\alpha n_\nu.
  \end{eqnarray}
  Any tensor may be projected onto $\Sigma_t$ by using the projection tensor
  ${h_\mu}^\nu$.  In this way,
  we can define the induced covariant derivative on $\Sigma_t$ as the projection
  (of every index) of the covariant derivative of the tensor in $\cal M$.
  For example,
 \begin{equation}
   D_\mu T^{\nu\rho} \equiv {h_\mu}^\alpha {h^\nu}_\beta {h^\rho}_\gamma 
      \nabla_\alpha T^{\beta\gamma}.
 \end{equation}
  where $T^{\nu\rho}$ is a tensor field defined on $\Sigma_t$.
  We now want to write the curvature scalar for $\cal M$ in terms of quantities 
  defined on $\Sigma_t$.  We begin by decomposing it in terms of the Einstein 
  and Ricci tensors,
  \begin{equation}
   \label{eqn:curvature_decomposition}
   R = 2(G_{\mu\nu} - R_{\mu\nu})n^\mu n^\nu.
  \end{equation}
  From the Gauss-Codazzi relations,  we obtain the initial value constraint
  \begin{equation}
   \label{eqn:gauss-codacci}
   G_{\mu\nu}n^\mu n^\nu = \frac{1}{2}({\cal R} - K_{\mu\nu}K^{\mu\nu} + K^2).
  \end{equation}
  By definition, the Riemann tensor satisfies
 \begin{equation}
   R_{\mu\nu}n^\mu n^\nu = -n^\mu (\nabla_\mu \nabla_\nu - 
    \nabla_\nu \nabla_\mu ) n^\nu,
 \end{equation}
  which, after some simplification, yields
  \begin{equation}
   \label{eqn:ricci}
   R_{\mu\nu}n^\mu n^\nu = K^2 - K_{\mu\nu}K^{\mu\nu} - 
    \nabla_\mu(n^\mu K - a^\mu ),
  \end{equation}
  where $a^\mu = n^\nu \nabla_\nu n^\mu$ is the acceleration of the unit 
  normal $n^\mu$.
  If we substitute equations (\ref{eqn:gauss-codacci}) and (\ref{eqn:ricci})
  into equation (\ref{eqn:curvature_decomposition}), we obtain the desired 
  expression for $R$, 
  \begin{equation}
   \label{eqn:curvature}
   R = {\cal R} + K_{\mu\nu}K^{\mu\nu} - K^2 + 2 \nabla_\mu(n^\mu K - a^\mu ).
  \end{equation}
  We now want to write this expression in terms of the canonical variables,
  $P^{\mu\nu}$, $h_{\mu\nu}$, $N$, and $N_\mu$. 
  We begin by expressing the extrinsic curvature in terms of these canonical 
  variables,
 \begin{equation}
   \label{eqn:momentum}
   K_{\mu\nu} = \frac{1}{2N}(\dot{h}_{\mu\nu} - 2D_{(\mu}N_{\nu)} ),
 \end{equation}
  where $\dot{h}_{\mu\nu}$ indicates the Lie derivative along the 
  evolution vector $t^\mu$.  If we substitute this into the action, we can 
  calculate the momentum conjugate to the hypersurface metric $h_{\mu\nu}$,
  \begin{equation}
   \label{eqn:mom}
   P^{\mu\nu} = \frac{1}{2\kappa}\sqrt{h}(K^{\mu\nu}- Kh^{\mu\nu}).
  \end{equation}
  If we calculate the term due to the extrinsic curvature in equation 
  (\ref{eqn:curvature}), 
 \begin{equation}
   K_{\mu\nu} K^{\mu\nu} - K^2 = \frac{2\kappa}{N\sqrt{h}}
    [P^{\mu\nu}\dot{h}_{\mu\nu} - \frac{\kappa N}{\sqrt{h}}
    (2P_{\mu\nu}P^{\mu\nu} - P^2) - 2P^{\mu\nu}D_\mu N_\nu ], 
 \end{equation}
  we can then define the energy constraint as the term in the action which 
  vanishes when the variation of $N$ is set to zero, 
 \begin{equation}
   {\cal H} = \frac{\kappa}{\sqrt{h}}(2P_{\mu\nu}P^{\mu\nu} - P^2) -
	      \frac{\sqrt{h}}{2\kappa} {\cal R},
 \end{equation}
  and hence we see that 
  \begin{equation}
   \label{eqn:integrand}
   \frac{1}{2\kappa}N\sqrt{h}[{\cal R} + K_{\mu\nu} K^{\mu\nu} - K^2] = 
    P^{\mu\nu}\dot{h}_{\mu\nu} - N{\cal H} - 2P^{\mu\nu}D_\mu N_\nu .
  \end{equation}
 If we write   
 \begin{equation}
  P^{\mu\nu}D_\mu N_\nu  = D_\mu (P^{\mu\nu}N_\nu ) - N_\mu D_\nu P^{\mu\nu},
 \end{equation}
 then we can define the momentum constraint (which vanishes when the variation
 due to the shift vector vanishes) as 
 \begin{equation}
  {\cal H^\mu} = - D_\nu P^{\mu\nu},
 \end{equation}
 and hence we obtain 
 \begin{equation}
  \label{eqn:mom_constraint}
  P^{\mu\nu}D_\mu N_\nu  = D_\mu (P^{\mu\nu}N_\nu ) + N^\mu {\cal H}_\mu.
 \end{equation}

 \subsection{The three-boundary $\cal B$}

  The induced metric and extrinsic curvature of $\cal B$ are given by
  \begin{eqnarray}
   \gamma_{\mu\nu} & = & g_{\mu\nu} - u_\mu u_\nu, \\
   \Theta_{\mu\nu} & = & {\gamma_\mu}^\alpha \nabla_\alpha u_\nu.
  \end{eqnarray}
  Since we are not assuming that $n^\mu$ is in the tangent space to $\cal B$,
  the scalar product of the two,
  \begin{equation}
   \eta = n_\mu u^\mu,
  \end{equation}
  may be nonzero.
  Moreover, if $\eta$ is nonvanishing, then the normalized restriction of 
  $n^\mu$ to ${\cal B}$,
  \begin{equation}
   \label{eqn:vmu}
   v^\mu = \lambda{\gamma^\mu}_\alpha n^\alpha = \lambda(n^\mu - \eta u^\mu),
  \end{equation} 
  will not, in general, be equal to $n^\mu$.  If we want $v^\mu$ to be a unit 
  timelike vector, we find that
  \begin{equation}
   \label{eqn:lambda}
   \lambda = \frac{1}{\sqrt{1+\eta^2}}.
  \end{equation}
 
 \subsection{The family of two-surfaces $B_t$}

  We want to consider $B_t$ as a hypersurface embedded in $\Sigma_t$.  Thus, 
  the
  normal to $B_t$ must lie in the tangent plane to $\Sigma_t$.  Therefore, 
  if $u^\mu$ is
  not orthogonal to $n^\mu$ on $\cal B$, then we cannot take $u^\mu$ to be the 
  normal to $B_t$.  Instead, we must take the normalized restriction of 
  $u^\mu$ to $\Sigma_t$, 
 \begin{equation}
   r_\mu = \lambda {h_\mu}^{\alpha} u_\alpha = \lambda(u_\mu + \eta n_\mu ),
 \end{equation}
 where the normalization constant is again $\lambda$, as given by equation
 (\ref{eqn:lambda}). 
 
  Thus, the induced metric is given by
 \begin{equation}
   \sigma_{\mu\nu} = h_{\mu\nu} - r_\mu r_\nu
      =  g_{\mu\nu} + \lambda^2(n_\mu n_\nu - u_\mu u_\nu 
	      - 2\eta n_{(\mu}u_{\nu )} ).
 \end{equation}

  We now want to express the extrinsic curvature of $B_t$ in terms of
  quantities defined on $\Sigma_t$ and $\cal B$.  By definition, it is 
 \begin{equation}
   k_{\mu\nu} = {\sigma_\mu}^\alpha D_\alpha r_\nu . 
 \end{equation}
  If we expand ${\sigma_\mu}^\alpha D_\alpha$, and take the trace, we obtain 
 \begin{equation}
   k = \nabla_\mu r^\mu + n^\mu n^\nu \nabla_\nu r_\mu.
 \end{equation}
  But, using the orthogonality of $n_\mu$ and $r_\mu$, we see that
 \begin{equation}
   n^\mu n^\nu \nabla_\nu r_\mu = -r_\mu n^\nu \nabla_\nu n^\mu 
    = -\lambda u_\nu a^\nu, 
 \end{equation}
  and by the definition of the extrinsic curvature of the $\Sigma_t$ and
  $\cal B$,
 \begin{equation}
   \nabla_\mu r^\mu = \lambda[\Theta + \eta K + \lambda^2(n^\mu - \eta u^\mu )
    \nabla_\mu \eta ].
 \end{equation}
  Combining the two results, and using the definition of the restriction of 
  $n^\mu$ 
  to $B_t$ from equation (\ref{eqn:vmu}), we obtain the following value for $k$,
  \begin{equation}
   \label{eqn:two_curvature}
    k = 
    \lambda [\Theta + \eta K - u_\mu a^\mu + \lambda v^\mu
     \nabla_\mu \eta ]. 
  \end{equation}         
  Note that since $v^\mu$ is the projection of $n^\mu$ onto $\cal B$, the
  derivative in the expression for $k$ is in the direction perpendicular to
  $B_t$ which lies in $\cal B$. 

  If we consider the momentum tensor density conjugate to $h_{ij}$, 
  as given by equation
  (\ref{eqn:mom}), then it takes a different value when viewed as a density on 
  $B_t$, because the volume element has changed.  On $B_t$ it is given by
  \begin{equation}
   \label{eqn:two_mom}
   P_\sigma^{\mu\nu} = \frac{1}{2\kappa}\sqrt{\sigma}(K^{\mu\nu} - Kh^{\mu\nu}).
  \end{equation}

 \section{Coordinate Conditions}

  We now present some formulae due to the particular coordinate system chosen,
  that is, oriented with respect to the hypersurfaces $\Sigma_t$, and the 
  three-boundary ${\cal B}$.  The formulae are primarily aimed at calculating
  the volume elements in terms of each other. 
  In our chosen coordinate system, we see that the unit normals are given by
  \begin{eqnarray}
   u_\mu & = & \left( -N, 0, 0, 0 \right), \\
   n_\mu & = & \left( 0, \sqrt{g^{11}}, 0, 0 \right). 
  \end{eqnarray}
  If we then calculate $\eta$, we see that
 \begin{equation}
   \eta = g^{\mu\nu}n_\mu u_\nu = -g^{01}N\sqrt{g^{11}}.
 \end{equation}
  We can use this expression to relate the lapse function and the intersection 
  variable to 
  the first component of the vector $u^\mu$, 
 \begin{equation}
   u^0 = g^{0\mu}u_\mu = g^{01}\sqrt{g^{11}} = -\frac{\eta}{N}.
 \end{equation}
 
  To calculate the relationships between the determinants of the metrics, we
  use the matrix identity,
 \begin{equation}
  \label{eqn:matrix}
   (A^{-1})_{ij} = \frac{1}{\det A}(-1)^{i+j}\det A(i,j),
 \end{equation}
  where $A(i,j)$ is the matrix formed by removing the $i^{\rm th}$ row and
  $j^{\rm th}$ column from $A$.  Then, since $g_{ij}(0,0) = h_{ij}$, 
  we see that
 \begin{equation}
   g^{00} = \frac{1}{g}h.
 \end{equation}
  But, $g^{00} = -N^{-2}$, and hence we obtain the familiar relation
 \begin{equation}
   \sqrt{-g} = N\sqrt{h}.
 \end{equation}
  Using the definition $\gamma_{ab}(0,0) = \sigma_{ab}$, equation 
  (\ref{eqn:matrix}) yields
 \begin{equation}
   \gamma^{00} = \frac{1}{\gamma}\sigma,
 \end{equation}
  If we then substitute in the value of $\gamma^{00}$,
 \begin{equation}
   \gamma^{00} = g^{00} - u^0 u^0 = -(\frac{1}{N^2} + \frac{\eta^2}{N^2})
          = - \frac{1}{N^2}(1+\eta^2) = -\frac{1}{N^2\lambda^2}, 
 \end{equation}
  we obtain the desired relation,
  \begin{equation}
   \label{eqn:gamma_sigma}
     \sqrt{-\gamma} = N\lambda\sqrt{\sigma}.
  \end{equation}
 
\section{Euclidean Formulae}

There are several important definitions which are changed when we deal with a
Euclidean rather than a Lorentzian metric.  The method of obtaining the
Hamiltonian from the action remains the same, but
several of the final results contain negative signs relative to their 
Lorentzian counterparts.  Euclidean quantities, such as the 
action and Hamiltonian, will be denoted by a circumflex.  The Hilbert 
action for Euclidean general relativity is  
\begin{eqnarray}
 \hat{I} & = & -\frac{1}{16\pi} \int_{\cal M} \sqrt{g} R - 
            \frac{1}{8\pi} \int_{\Sigma_0}^{\Sigma_1} \sqrt{h} K -
            \frac{1}{8\pi} \int_{\cal B} \sqrt{\gamma} \Theta \nonumber \\
 & & - \int_{B_0}^{B_1} d^2x\, \sqrt{\sigma} \cos^{-1} \eta. 
\end{eqnarray} 
Performing the same steps in decomposing the action as were followed in the 
Lorentzian case, we obtain
\begin{equation}
 \hat{I} = \int dt \left\{ \int_{\Sigma_t} d^3x\,P^{\mu\nu}\dot{h}_{\mu\nu} + 
       (\hat{H}_c + \hat{H}_k + \hat{H}_t + \hat{H}_m)
      \right\}, 
 \end{equation}
 where the Hamiltonian, 
 \begin{equation}
  \hat{H}[{\cal M}, g] = \hat{H}_c + \hat{H}_k + \hat{H}_t + \hat{H}_m, 
 \end{equation}
 is again a sum of four distinct terms:
 \begin{enumerate}
  \item a constraint term, 
   \begin{equation}
    \hat{H}_c = - \int_{\Sigma_t} d^3x\,[ N{\cal H} + N_\mu{\cal H}^\mu ],
   \end{equation}
  \item a curvature term, 
   \begin{equation}
    \hat{H}_k =  - \frac{1}{\kappa} \int_{B_t} d^2x\,N\sqrt{\sigma}k,
   \end{equation}
  \item a tilting term, 
   \begin{equation}
    \hat{H}_t = - \frac{1}{\kappa} \int_{B_t} d^2x\,N\lambda \sqrt{\sigma}
       \cos^{-1} \eta \nabla_\mu v^\mu , 
   \end{equation}
  \item and a momentum term, 
   \begin{equation}
    \hat{H}_m = - 2 \int_{B_t} d^2x\,r_\mu P_{\sigma}^{\mu\nu}N_\nu . 
   \end{equation}
 \end{enumerate}

In order to derive these results, it is necessary to note that $n^\mu$ is now
a spacelike vector, and hence it induces a metric
\[
 h_{\mu\nu} = g_{\mu\nu} - n_\mu n_\nu.
\]
The normal to the two-boundary is then
\[
 r_\mu = \lambda(u_\mu - \eta n_\mu),
\]
where the normalization constant $\lambda$ is now 
\[
 \lambda = \frac{1}{\sqrt{1-\eta^2}}.
\]
Apart from these changes, the analysis is identical.

\end{document}